\documentclass[lettersize,journal]{IEEEtran}
\usepackage{amsmath,amsfonts}
\usepackage{algorithmic}
\usepackage{algorithm}
\usepackage{array}
\usepackage[caption=false,font=normalsize,labelfont=sf,textfont=sf]{subfig}
\usepackage{textcomp}
\usepackage{stfloats}
\usepackage{url}
\usepackage{verbatim}
\usepackage{graphicx}
\usepackage{cite}
\usepackage{multirow}
\usepackage{balance}
\usepackage{flushend}
\usepackage{pbalance}

\def\BibTeX{{\rm B\kern-.05em{\sc i\kern-.025em b}\kern-.08em
    T\kern-.1667em\lower.7ex\hbox{E}\kern-.125emX}}
    
\usepackage{tikz}
\usepackage{array}
\newcommand*\filledcircled[1]{\tikz[baseline=(char.base)]{
    \node[shape=circle, draw, inner sep=0.25pt, fill=black, text=white] (char) {#1};}}

\begin{document}

\title{M3D-FLOW: A Monolithic 3-D-Stackable Processing-Near-DRAM Architecture for Generative Normalizing Flow Applications}

\author{Bipin Thapa Magar,~\IEEEmembership{Graduate Student Member,~IEEE,} Ishan Thakkar,~\IEEEmembership{Member,~IEEE,} \\and Sudeep Pasricha,~\IEEEmembership{Fellow,~IEEE}%
\thanks{Bipin Thapa Magar and Ishan Thakkar are with the Department of Electrical and Computer Engineering, University of Kentucky, Lexington, KY 40506, USA (e-mail: bipin.thapa@uky.edu; igthakkar@uky.edu).}%
\thanks{Sudeep Pasricha is with the Department of Electrical and Computer Engineering, Colorado State University, Fort Collins, CO 80523, USA (e-mail: sudeep@colostate.edu).}%
}

\markboth{Accepted in IEEE Transactions on Computer-Aided Design of Integrated Circuits and Systems}%
{Thapa Magar \MakeLowercase{\textit{et al.}}: [M3D-FLOW: A Monolithic 3-D-Stackable Processing-Near-DRAM Architecture for Generative Normalizing Flow Applications]}
\IEEEpubid{\begin{minipage}{\textwidth}
\centering
\scriptsize
This work has been accepted for publication in IEEE Transactions on Computer-Aided Design of Integrated Circuits and Systems.\\
\copyright~2026 IEEE. Personal use of this material is permitted. Permission from IEEE must be obtained for all other uses, in any current or future media, including reprinting/republishing this material for advertising or promotional purposes, creating new collective works, for resale or redistribution to servers or lists, or reuse of any copyrighted component of this work in other works.
\end{minipage}}

\maketitle

\begin{abstract}
Normalizing Flows (NFs) have shown strong results in generative modeling, recently reaching state-of-the-art image generation performance. However, their repetitive, sequential operations limit the computational and memory efficiency achievable with GPUs and TPUs. We propose M3D-FLOW, a novel monolithic 3D-stackable (M3D) DRAM-based processing-near-memory (PNM) architecture that eliminates memory bottlenecks using a Cu-Cu hybrid-bonded logic die integrated with an M3D DRAM die. The M3D DRAM provides dense bank integration and a tiered organization with variable access latencies that M3D-FLOW exploits for the efficient processing of different NF workloads. To maximize compute density while minimizing area, energy, and thermal overhead, M3D-FLOW employs hybrid binary--stochastic floating-point multipliers, performing mantissa multiplication in the stochastic domain and exponent operations in the binary domain. A pipelined execution strategy maps sequential NF stages across dynamically partitioned PE groups, and together with the stochastic hardware, enables M3D-FLOW to achieve at least $6.6\times$ higher performance and $3\times$ better energy efficiency than prior PNM accelerators, with only a 2.2\% bits-per-dimension (BPD) increase over bfloat16 and 4.1\% over FP32. An analysis of stochastic hardware variants further shows that shorter stochastic bitstreams reduce execution time by up to $3.3\times$ at the cost of higher approximation error, providing a concrete design guide for latency--accuracy trade-offs. 

\end{abstract}

\begin{IEEEkeywords}
Monolithic 3-D (M3D) DRAM, normalizing flows (NFs), processing near memory (PNM), stochastic computing (SC)
\end{IEEEkeywords}

\section{Introduction}
\IEEEPARstart{T}{he} rapid progress of generative modeling has enabled state-of-the-art performance in image synthesis, density estimation, and representation learning \cite{ho2020denoising,brown2020gpt3,tarflow}. Among modern generative models, Normalizing Flows (NFs) have emerged as a powerful class of architectures that learn expressive probability distributions while supporting exact likelihood evaluation \cite{rezende2015variational,nice,realnvp,glow,maf,tarflow}. By composing a sequence of invertible transformations, NFs map complex data distributions into tractable latent spaces, enabling efficient training and inference \cite{rezende2015variational}. Recent NF architectures \cite{maf,tarflow,kingma2016iaf} further improve expressiveness and sample quality, making NFs competitive with diffusion-based and other generative models.

Despite these advances, NF inference is fundamentally memory-bound. Unlike conventional feedforward networks, NFs apply many sequential transformations, resulting in frequent transfers of large weight tensors and intermediate activations between memory and compute units \cite{rezende2015variational,kingma2016iaf}. Core NF operations, including matrix multiplications, point-wise multiplications, additions, and nonlinear transformations, repeatedly switch between computation types and touch large parameter sets. This pattern leads to frequent off-chip DRAM accesses and underutilized compute units on conventional GPU and TPU architectures \cite{huang2019gpu}. Therefore, NF workloads are often limited by memory bandwidth and data movement overheads rather than raw floating-point operation (FLOP) capability.

Processing-near-memory (PNM) architectures address these bottlenecks by reducing data movement and executing computation closer to memory \cite{mutlu2019processing,mutlu2019enabling,ghose2019pim}. By exploiting the high internal bandwidth of DRAM and integrating compute logic near memory, PNM designs \cite{kim2018grim,boroumand2021polynesia,
besta2021sisa,singh2020nero,hashemi2016accelerating} can significantly lower the latency and energy cost of data transfers. However, existing PNM solutions are constrained by limited compute density, modest bandwidth scalability, and stringent thermal envelopes. These constraints become particularly pronounced for NF workloads, which demand sustained high-throughput computation across tens to hundreds of sequential stages \cite{tarflow}.
\IEEEpubidadjcol
Recent advances in monolithic 3D-stackable DRAM (M3D) and hybrid bonding introduce new opportunities to overcome these limitations. M3D integration enables vertically stacked DRAM arrays with extremely high density \cite{han2023ongoing,hsu2025monolithic,hsu2025monolithicbeol}, while hybrid bonding provides fine-pitch, low-latency Cu–Cu interconnects between memory and logic dies \cite{chen20203d,fujun2020stacked,qpshybrid,choi2024three}. Together, these technologies support tightly coupled memory–compute systems with massive internal bandwidth, which are well-suited to the bandwidth-hungry nature of NF workloads. At the same time, placing logic directly beneath stacked DRAM introduces significant design challenges: power density and heat flow are tightly coupled to DRAM reliability, logic area is constrained by the DRAM footprint, and computation must be highly energy-efficient to remain within safe operating temperatures \cite{han2021power,wu2025signal}.

In this work, we propose M3D-FLOW, a novel monolithic 3D-stackable DRAM-based PNM architecture tailored to NF workloads. M3D-FLOW leverages hybrid-bonded M3D DRAM to provide high-bandwidth, low-latency access to model parameters and activations, while dedicated processing engines (PEs) located beneath DRAM banks execute NF computations near memory. To maximize compute density under tight thermal constraints, M3D-FLOW employs a hybrid binary–stochastic floating-point arithmetic: mantissa multiplication is performed in the stochastic domain, while exponent operations remain in the binary domain, substantially reducing area and energy with minimal loss in numerical accuracy. To fully exploit the sequential structure of NFs, M3D-FLOW introduces a pipelined execution strategy that maps different flow stages across dynamically partitioned PE groups, enabling concurrent execution of multiple stages and improving hardware utilization. Finally, an optimized dataflow and prefetching mechanism overlaps DRAM access with computation, hiding memory latency and saturating the high internal bandwidth of the M3D DRAM stack.

The key contributions of this work are:

\begin{itemize}
    \item We introduce M3D-FLOW, a novel hybrid-bonded monolithic 3D-stackable PNM architecture designed specifically to accelerate generative NF applications.
    \item We propose a hybrid floating-point design that performs mantissa multiplication stochastically while retaining exponent operations in binary, achieving significant reductions in area and power with minor accuracy loss.
    \item We demonstrate that integrating low-power stochastic computing within a hybrid-bonded M3D stack enables higher multiply-accumulate (MAC) density while maintaining safe DRAM operating temperatures, addressing key thermal challenges in 3D-stacked systems.
    \item We design a dataflow that overlaps DRAM access with computation via SRAM buffering and prefetching, and a pipelined execution strategy that maps sequential NF stages across dynamically partitioned PE sets to enable pipeline parallelism and higher throughput.
    \item We explore multiple stochastic multiplier configurations and analyze latency–area–accuracy trade-offs, identifying practical operating points for different system constraints.
    \item We benchmark M3D-FLOW in performance and energy efficiency across multiple NF models against state-of-the-art CPU, GPU, TPU, and prior PNM accelerators.
\end{itemize}

\section{Background}
 
\subsection{Normalizing Flows}
 
NFs transform a simple base density $p_z(\mathbf{z})$, typically a standard Gaussian, through invertible and differentiable mappings $f$ \cite{rezende2015variational}. For $\mathbf{x}=f(\mathbf{z})$, the exact log-likelihood follows the change-of-variables formula:
 
\begin{equation}
    \label{eq:nf}
    \log p_x(\mathbf{x}) = \log p_z(f^{-1}(\mathbf{x})) +
    \log \left| \det \left( \frac{\partial f^{-1}}{\partial \mathbf{x}}
    \right) \right|
\end{equation}
 
Composing mappings $f_k$ transforms the base distribution into a complex target distribution. Tractable flows commonly use triangular Jacobians, reducing the determinant to a product of diagonal elements. NICE introduced additive coupling \cite{nice}; RealNVP added learned affine scaling \cite{realnvp}; MAF enabled exact density evaluation using masked autoregressive models \cite{maf}; Glow introduced learnable $1\times1$ invertible convolutions \cite{glow}; and TarFlow incorporated masked transformer blocks for high-quality likelihood estimation \cite{tarflow}.
 
\subsection{Stochastic Computing}
 
SC represents values through bitstream probabilities rather than binary encodings. An $n$-bit value is represented by a length-$2^n$ bitstream whose fraction of 1s encodes the value. Bitstreams can be rate-coded, with randomly distributed 1s \cite{wu2020ugemm}, or deterministic, such as transition-coded unary (TCU), where 1s are grouped at one end \cite{vatsavai2023multiplier}. Unipolar and bipolar encodings represent $[0,1]$ and $[-1,1]$, respectively, using stochastic number generators (SNGs) \cite{kim2016rng,scope,vatsavai2023multiplier}.
 
SC simplifies arithmetic: multiplication uses an AND gate for unipolar encoding or an XNOR gate for bipolar encoding, while addition uses multiplexers or counters \cite{ren2017scdcnn}. Prior work has improved SC matrix multiplication through conditional probability \cite{wu2020ugemm} and implemented nonlinear functions, pooling, and activations \cite{ren2017scdcnn,li2018heif,li2023hybrid}. These properties make SC suitable for energy-efficient acceleration of error-tolerant workloads such as deep neural networks \cite{alaghi2013survey}.
 
Its main limitations are correlation error and the long bitstreams needed for high precision. Error-free multiplication of $n$-bit values theoretically requires $2^{2n}$ bits, increasing storage and energy cost. Hierarchical and hybrid encodings reduce this overhead \cite{scope,alaghi2013survey,li2023hybrid}. Stochastic-to-binary conversion commonly uses popcount operations and may introduce precision loss \cite{atria}; delayed accumulation \cite{scope} and in-memory analog accumulation with ADC-based conversion \cite{agni,star,artemis} reduce this cost.
 
\subsection{Monolithic 3D-Stackable DRAM and Hybrid Bonding}
 
M3D-Stackable DRAM extends DRAM scaling beyond sub-10-nm nodes without the complexity and cost of TSV-based stacks such as HBM \cite{han2023ongoing,hsu2025monolithic,hsu2025monolithicbeol}. M3D sequentially fabricates DRAM layers on the same wafer \cite{han2023ongoing,choi2024three,hsu2025monolithic}, using vertically stacked horizontal 1T1C cells connected by vertical bitlines and wordline staircases. Eliminating TSVs enables thinner dies, improved vertical thermal conduction, and high memory density.
 
HB connects the M3D stack to logic through direct Cu--Cu fusion \cite{chen20203d,fujun2020stacked,qpshybrid}, allowing independent fabrication and face-to-face bonding at each die's optimal process node \cite{fujun2020stacked,3dpath}. With 1--3~$\mu$m pitch and up to 110,000 vias/mm$^2$ \cite{h2llm,fujun2020stacked,stratum,cao2026two}, HB is much denser than TSVs or microbumps, bypasses external I/O limits \cite{stratum}, and enables tens of TB/s internal bandwidth \cite{stratum}. Its short Cu--Cu connections also reduce parasitic resistance and capacitance relative to TSVs \cite{3dpath}, achieving I/O energy as low as 0.3--0.88~pJ/bit \cite{h2llm,qpshybrid}.
 
\subsection{Processing-In-Memory}
 
The separation of processors and memory causes the memory wall, where data movement dominates execution time and energy \cite{pandiyan2014energy,kanev2015profiling,mutlu2019processing,mutlu2019enabling,ghose2019pim}. Processing-In-Memory (PIM) addresses this challenge by moving computation closer to memory \cite{mutlu2022modern}.
 
PIM includes two main classes \cite{mutlu2022modern}. Processing-using-Memory (PUM) exploits memory-array device properties to perform row-granularity bulk operations, including bitwise logic and simple arithmetic, with minimal data movement \cite{scope,atria,agni,star,artemis,heo2024neupims,magar2026dristi,ambit,roc}. Processing-near-Memory (PNM) places dedicated logic near DRAM banks \cite{h2llm,stratum,yue2024exploiting,cao2026two}, in 3D-stacked logic layers \cite{kim2018grim,besta2021sisa}, or in memory controllers \cite{singh2020nero,hashemi2016accelerating}. This flexibility supports more complex operations and diverse workloads, including graph processing, machine learning, and database analytics \cite{nai2017graphpim,boroumand2018google}. Hybrid-bonded M3D PNM designs further integrate high-bandwidth logic beneath DRAM to improve performance and energy efficiency \cite{3dpath,h2llm,stratum,yue2024exploiting,cao2026two}.

\begin{figure}[t]
    \centering
    \includegraphics[width=\columnwidth]{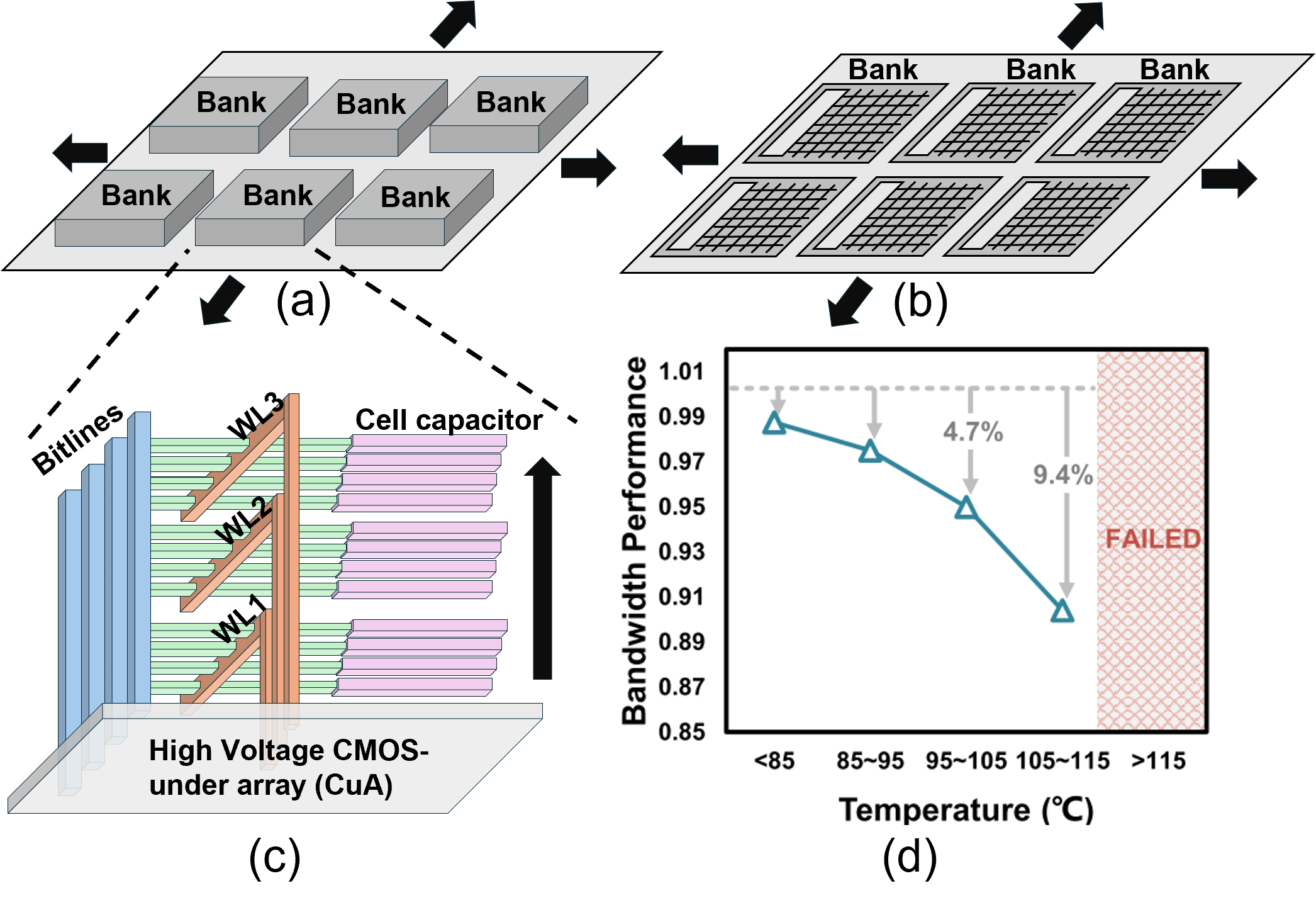}
    \caption{(a) Monolithic 3D-Stackable (M3D) DRAM, (b) Conventional 2D DRAM, and (c) the bank structure of a vertically stacked horizontal 1T1C DRAM. The arrows indicate directions for increased capacity: 2D DRAM capacity increases primarily with area, while M3D can scale across both area and height (stacking density). (d) Bandwidth performance of DRAM vs temperature \cite{3dpath}. WL: Wordline.}
    \label{fig:2dvs3d}
\end{figure}

\section{M3D-FLOW: Motivation and Novelty} \label{sec:motivation}
 
\subsection{Challenges with NF Acceleration}
To the best of our knowledge, M3D-FLOW is the first dedicated accelerator for normalizing flow inference. NF acceleration is challenging because, unlike diffusion models that reuse a denoising network across timesteps or autoregressive models that reuse a fixed network across tokens, each flow stage has a distinct weight tensor. Sequential stage dependencies prevent the data reuse and batching that conventional accelerators use to amortize memory accesses \cite{tarflow,glow}. Each stage streams a fresh weight tensor through compute once per forward pass, making memory bandwidth, rather than arithmetic throughput, the primary bottleneck. GPU execution is therefore dominated by repeated off-chip DRAM accesses, motivating near-memory execution that keeps weights and activations close to compute.
 
\subsection{Motivation for Hybrid-Bonded M3D Stacking}
Conventional 2D DRAM scaling faces physical limits below 10~nm \cite{hsu2025monolithic}. Although High Bandwidth Memory (HBM) increases bandwidth through die stacking, TSVs add area overhead and limit vertical connection density, constraining near-memory processing \cite{stratum}. Monolithic 3D (M3D) integration instead provides fine-grained vertical scalability \cite{han2023ongoing,hsu2025monolithic}. As shown in Fig.~\ref{fig:2dvs3d}(a)-(c), M3D replaces lateral DRAM-bank placement with vertically stacked layers, increasing capacity without planar footprint growth. M3D can provide up to $5.2\times$ higher memory density than DDR5 \cite{stratum,choi2024ddr5}, helping retain large NF models, such as TarFlow with over 300M parameters \cite{tarflow}, on chip.
 
HB connects DRAM and logic dies through direct Cu--Cu fusion at $1~\mu$m pitch \cite{chen20203d}, approximately $5\times$ finer than HBM TSVs. It delivers 19--30~TB/s internal bandwidth \cite{stratum}, compared with about 800~GB/s for a single interposer-based HBM3 stack \cite{stratum,h2llm}. This more than $20\times$ bandwidth advantage is especially valuable for NFs, whose stages repeatedly stream distinct weight tensors. Recent HB-based PNM designs similarly exploit high internal bandwidth \cite{h2llm,stratum,3dpath,yue2024exploiting}.
 
\subsection{Motivation for Stochastic Computing}
Thermal management remains a key challenge in M3D DRAM. Heat from the logic die propagates upward into DRAM, degrading bandwidth above approximately 85°C (Fig.~\ref{fig:2dvs3d}(d)) because temperature increases leakage and refresh requirements \cite{3dpath,hsu2025monolithic}. This is particularly restrictive for NF acceleration, which requires many concurrent MAC units. Prior work addresses this using thermal-aware layouts \cite{3dpath}, reduced frequency \cite{h2llm}, or active cooling \cite{stratum,3dpath,han2021power}, often sacrificing throughput or increasing packaging complexity.
 
M3D-FLOW instead replaces binary mantissa multipliers with stochastic units. A stochastic mantissa multiplier uses bitwise AND operations rather than a binary multiplier tree, providing $13\times$ lower area and $46\times$ lower power. Exponent operations and addition remain binary because stochastic adders are $2.97\times$ larger and $1.46\times$ more power hungry. Despite conversion overhead, this hybrid approach increases MAC density while maintaining the logic-die thermal budget. These area and power savings also generalize beyond NF acceleration, benefiting area-constrained designs and thermally limited 3D-stacked systems.
 
\subsection{Novelty Relative to Prior Work}
M3D-FLOW combines several features that distinguish it from prior M3D PNM designs. Unlike Stratum \cite{stratum}, which shares a non-linear unit across Processing Engines (PEs) and lacks dedicated vector addition hardware, each M3D-FLOW PE integrates a non-linear engine (NLE), multiplier engine (ME), and vector adder engine (VAE). This eliminates inter-PE contention and removes element-wise additions from the MAC datapath.
 
Its hybrid binary-stochastic multiplier reduces per-MAC area, enabling higher MAC density within a fixed thermal budget, a capability absent from prior M3D architectures. M3D-FLOW also differs from stochastic in-DRAM designs \cite{scope,star,artemis}, which are restricted to array-level bitwise operations and cannot support the complete bfloat16 NF computation graph. Finally, its per-set weight assignment, tier-aware DRAM placement, and double-buffered prefetching are co-designed to reduce redundant DRAM accesses and hide remaining memory latency.
 
Together, M3D density, HB bandwidth, and stochastic power efficiency enable high-throughput, energy-efficient NF acceleration while maintaining safe DRAM temperatures.

\begin{figure*}[t]
    \centering
    \includegraphics[width=0.85\textwidth]{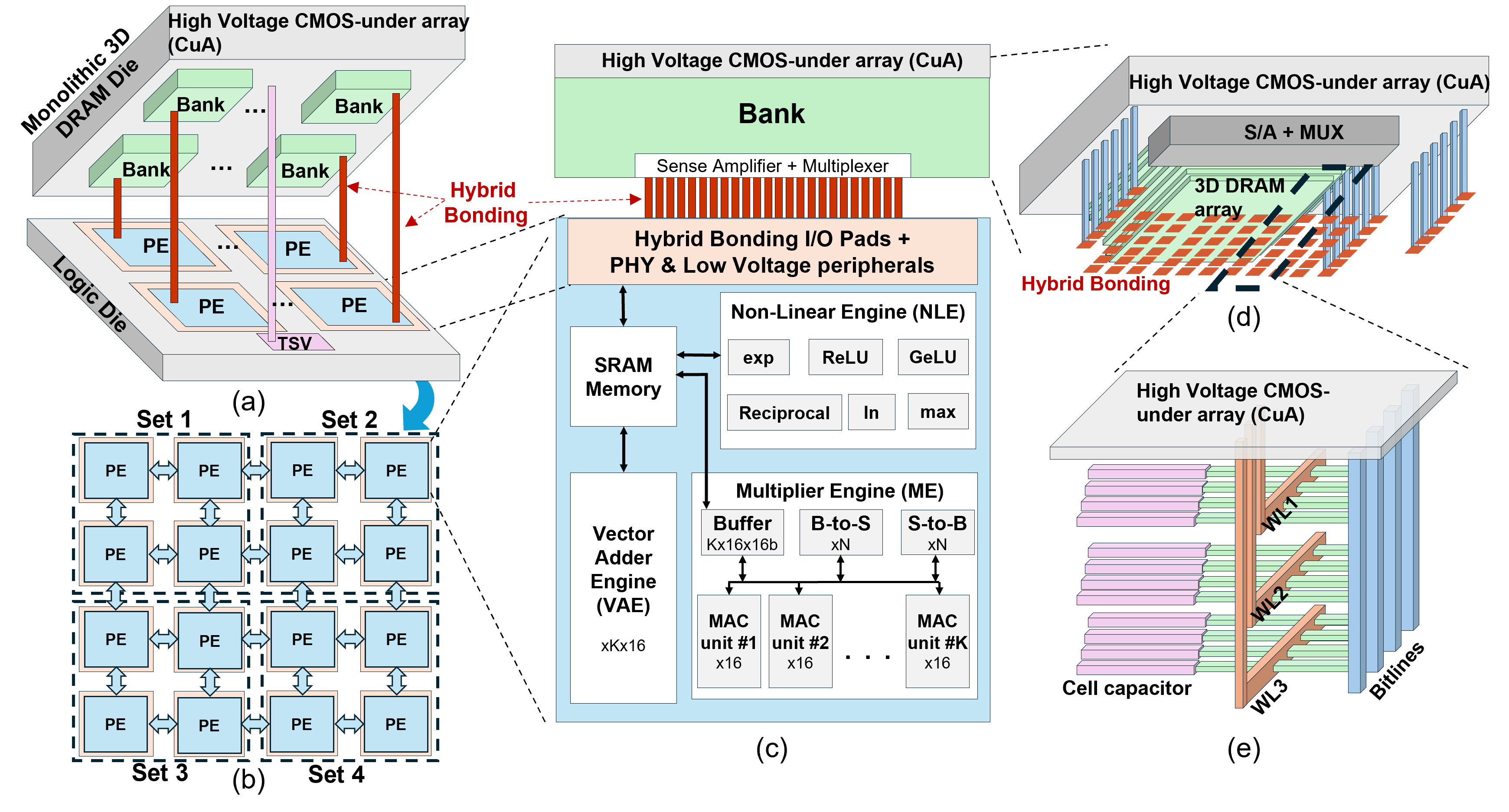}
    \caption{(a) M3D-FLOW architecture with hybrid bonded M3D DRAM (flipped) and logic dies, (b) example logic die with 16 PEs divided into four sets of 4 PEs each, (c)  M3D-FLOW PE architecture, (d) M3D DRAM bank with sense amplifier, multiplexer, hybrid bonding pads (flipped for hybrid bonding), and (e) structure of M3D DRAM array with vertically stacked horizontal 1T1C cells (flipped for hybrid bonding). B-to-S: Binary-to-Stochastic, S-to-B: Stochastic-to-Binary, MAC: Multiply-Accumulate, S/A: Sense Amplifiers, MUX: Multiplexer, WL: Wordline.}
    \label{fig:architecture}
\end{figure*}

\section{M3D-FLOW Architecture}

M3D-FLOW is built on a hybrid-bonded M3D stack (Fig.~\ref{fig:architecture}(a)) in which a logic die is tightly integrated beneath a DRAM die. Each DRAM bank is paired one-to-one with a dedicated Processing Engine (PE) on the logic die via Cu--Cu hybrid bonding, giving each PE direct, high-bandwidth access to its corresponding memory. As shown in Fig.~\ref{fig:architecture}(c)--(d), each PE comprises three specialized engines: the Multiplier Engine (ME), which performs hybrid binary-stochastic floating-point MAC operations; the Vector Adder Engine (VAE), which handles element-wise vector addition; and the Non-Linear Engine (NLE), which executes nonlinear functions via LUTs and dedicated logic. The following subsections describe each component of M3D-FLOW in detail.

\subsection{M3D DRAM and Hybrid Bonding} \label{sec:dram_arch}

M3D-FLOW's DRAM tier uses a monolithic 3D-stackable architecture with vertical bitlines and horizontal 1T1C cells \cite{hsu2025monolithic}, as illustrated in Fig.~\ref{fig:architecture}(a)--(b). 
 
High-voltage decoders are placed beneath the DRAM array via a CMOS-under-Array (CuA) process; they appear above the array because the DRAM die is flipped for hybrid bonding. Sense amplifiers (S/As) sit adjacent to the array, and transfer data to the logic die through vertical bitlines connected to the hybrid bonding (HB) I/O pads. Because the bitline pitch ($\approx$70~nm) is much finer than the HB pitch ($\approx$1~$\mu$m), a multiplexer alongside the S/As bridges this mismatch and simultaneously enables burst reads, reducing overall DRAM read latency.

A key structural feature of M3D is that wordlines (WLs) are routed through a staircase (highlighted in orange in Fig.~\ref{fig:architecture}(b)), which causes access latency to increase with layer depth. Rather than treating this as a pure liability, M3D-FLOW exploits the resulting latency gradient 
to form a tiered memory organization: 1024 layers are partitioned into eight tiers with $t_\text{RCD}$ values ranging from 2.29~ns to 22.88~ns \cite{stratum}.

By applying the dataflow mapping and execution strategies described in Section \ref{sec:optimization}, frequently accessed NF parameters are mapped to low-latency tiers, while less critical data occupies slower tiers, directly reducing average access latency. 

Additionally, hybrid bonding enables low-voltage memory peripherals, such as DQ buffers, level shifters, and address decoders, to be relocated from the DRAM die to the logic die, further increasing effective DRAM density.

\subsection{Processing Engine (PE)}
Each PE is the atomic compute unit of M3D-FLOW, paired directly with one DRAM bank via hybrid bonding for low-latency, high-bandwidth data access (Fig. \ref{fig:architecture}(a)). Each PE includes a local SRAM buffer to store model parameters 
and intermediate results, decoupling memory access from computation and enabling the prefetching strategy described in 
Section~\ref{sec:optimization}. PEs are also interconnected via a 
128~GB/s inter-PE link (prior works, e.g., \cite{stratum}, commonly use such bandwidths for inter-PE links), though this link is used exclusively during pipelined execution phases (Section~\ref{sec:pipelining}) to pass intermediate activations between flow stages. This link is implemented as a 1,024-bit point-to-point datapath at 1~GHz between adjacent PEs at set boundaries only, with no shared mesh, routing, or arbitration logic. Wide on-chip datapaths of this width have precedent in synthesized open-source interconnect implementations~\cite{kurth2021open}.
 
Each PE integrates three engines: the Multiplier Engine (ME) for matrix--matrix multiplication, the Vector Adder Engine (VAE) for element-wise vector additions, and the Non-Linear Engine (NLE) for nonlinear functions. The designs of these engines reflect the specific arithmetic demands of NF workloads, as discussed next.

\subsection{Multiplier Engine (ME)} \label{sec:me}

\begin{figure}[t]
    \centering
    \includegraphics[width=\columnwidth]{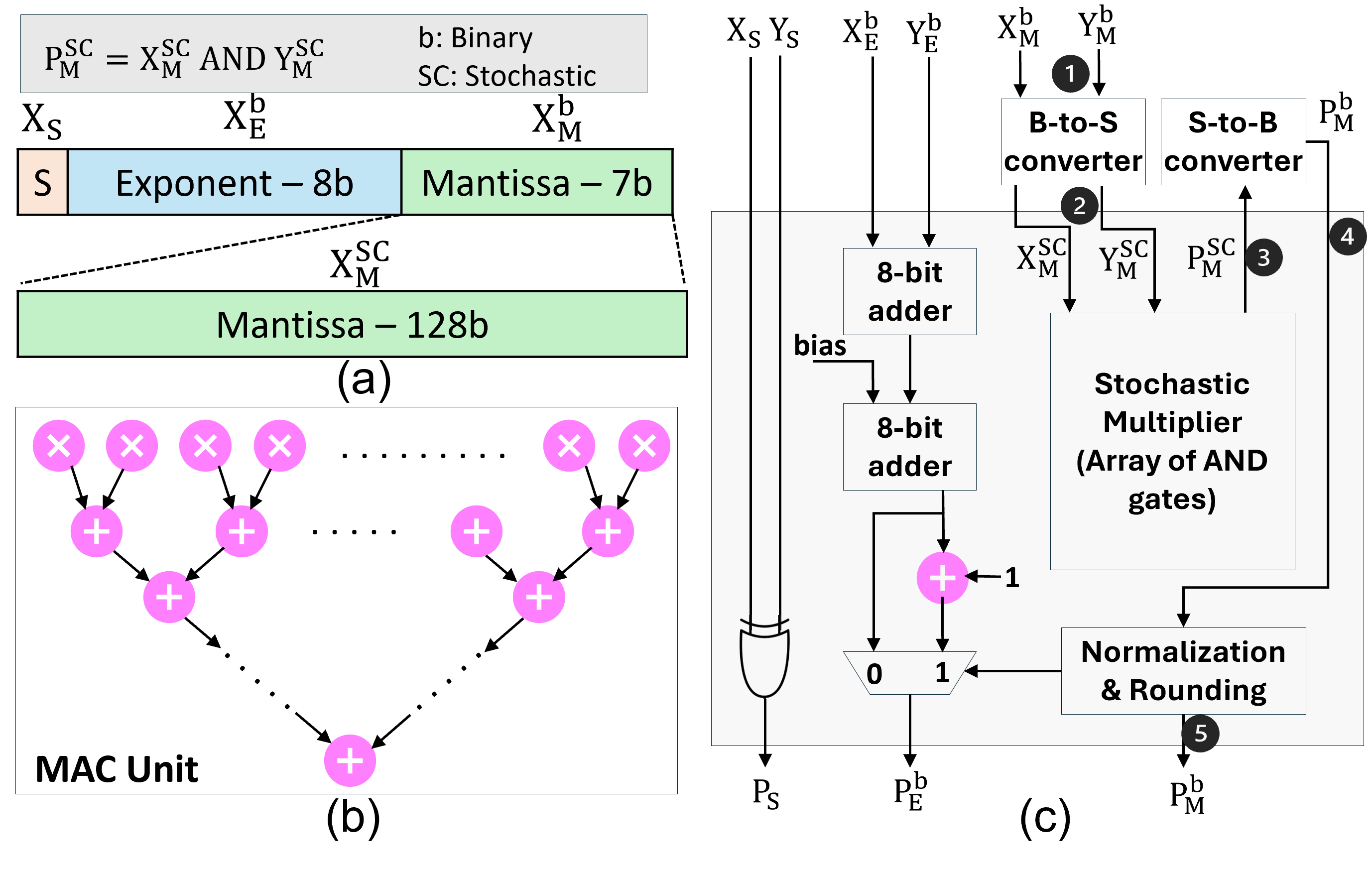}
    \caption{(a) bfloat16 (top) and stochastic (bottom) representations, (b) MAC unit structure with multipliers and adder trees, and (c) internal block diagram of the hybrid binary–stochastic multiplier. \protect\filledcircled{1}–\protect\filledcircled{5} indicate the stages of hybrid binary–stochastic mantissa multiplication. Here, $X$, $Y$ represent multiplicand and multiplier; $P$ denotes $X \times Y$, the resulting product.}
    \label{fig:multiply}
\end{figure}

The ME is the most area- and power-intensive component of each PE, and its design is central to M3D-FLOW's thermal efficiency. M3D-FLOW performs floating-point multiplication in bfloat16 format, where each value $X$ comprises a sign bit ($X_S$), an 8-bit exponent ($X_E^b$), and a 7-bit mantissa ($X_M^b$), as shown in Fig.~\ref{fig:multiply}(a). Multiplication proceeds in three stages: mantissa multiplication, exponent addition/subtraction, and normalization/rounding. Of these, mantissa multiplication dominates area and power, making it the natural target for stochastic optimization. Our evaluation confirms that stochastic mantissa multipliers require $13\times$ less area and $46\times$ less power than binary equivalents. Stochastic addition, in contrast, is less favorable because stochastic adders consume $2.97\times$ more area and $1.46\times$ more power than binary adders. Hence, exponent addition and normalization remain in the binary domain. M3D-FLOW therefore applies stochastic computing selectively: only the mantissa $X_M^b$ is converted to stochastic format $X_M^{SC}$ (Fig.~\ref{fig:multiply}(a)), while the 
sign and exponent are handled in binary throughout.

Binary-to-stochastic conversion is performed by $N$ B-to-S converters within each PE (Fig.~\ref{fig:architecture}(d)) using a deterministic scheme \cite{vatsavai2023multiplier,bresenham} that minimizes correlation-induced errors. The first operand is encoded in Transition-Coded Unary (TCU) format, grouping all logic-1s at the end of the bitstream \cite{wu2020ugemm,vatsavai2023multiplier}. Two options are supported for the second operand: a bit-position correlation (BPC) encoder \cite{vatsavai2023multiplier}, which generates fixed-length ($2^n$-bit) streams with lower hardware cost, or a precomputed lookup table (LUT) populated via the Bresenham algorithm \cite{bresenham}, which supports variable-length streams and can theoretically eliminate approximation error at $2^{2n}$ bits \cite{wu2020ugemm} at the cost of higher area. 
The area--latency--accuracy trade-offs between these two variants are analyzed in Section~\ref{sec:sensitivity}. Stochastic-to-binary 
conversion is performed by $N$ S-to-B converters using popcount-based adder trees.

The ME organizes its multipliers into $K$ MAC units, each containing 16 parallel multipliers followed by a reduction tree (Fig.~\ref{fig:multiply}(b)), for a total of $16 \times K$ MAC operations per PE per cycle. Since all $K$ MAC units share the $N$ conversion circuits, the ratio $N / (16 \times K)$ determines converter utilization: when $N < 16 \times K$, multipliers stall awaiting converters, making the joint selection of $K$ and $N$ a critical area-throughput design decision.

The internal datapath of a single multiplier is shown in 
Fig.~\ref{fig:multiply}(c). The sign bits $X_S$ and $Y_S$ are XORed to form $P_S$. The exponents $X_E^b$ and $Y_E^b$ are summed by an 8-bit binary adder, then bias-subtracted and normalized to yield $P_E^b$. For the mantissa, M3D\text{-}FLOW adopts a fully bit-parallel stochastic multiplier. The 7-bit mantissas $X_M^b$ and $Y_M^b$ are first converted in step~\filledcircled{1} to length-$L$ stochastic bitstreams $X_M^{SC}$ and $Y_M^{SC}$ with $L = 2^7 = 128$. In step~\filledcircled{2}, these streams are multiplied via 128 parallel bitwise AND operations to produce $P_M^{SC}$, so each mantissa product completes in a single clock cycle. Step~\filledcircled{3} converts $P_M^{SC}$ back to binary $P_M^b$, and steps~\filledcircled{4} and \filledcircled{5} normalize the result and combine it with the exponent. The stochastic unit (128-way AND array plus shared B-to-S and S-to-B converters) occupies $17~\mu\text{m}^2$ after synthesis, compared to $222~\mu\text{m}^2$ for a conventional binary mantissa multiplier tree, enabling both the reported $13\times$ area reduction and the 192~TFLOPS peak throughput under the bit-parallel design. Sign, exponent, 
and mantissa are then concatenated to form the final product. For standard MACs, 16 products feed the reduction tree; for element-wise multiplication, the reduction stage is bypassed. A dedicated ME buffer supports operand prefetching and intermediate storage throughout.

\subsection{Vector Adder Engine (VAE)}\label{sec:vae}
The VAE handles all element-wise floating-point additions required by NF models, and unlike the ME, it operates entirely in the binary domain. This design choice follows directly from the arithmetic analysis in Section~\ref{sec:me}: binary adders are $2.97\times$ smaller and $1.46\times$ more power-efficient than their stochastic counterparts, making the binary domain strictly preferable for addition. The VAE provides $K \times 16$ parallel FP adders, matching the ME's parallelism 
to avoid becoming a throughput bottleneck.

During operation, operands are first transferred from the PE’s SRAM into a dedicated VAE's local buffer of size $K \times 16 \times 16$ bits. The FP adders read directly from this buffer to perform element-wise additions in parallel, after which the results are written back to the PE’s SRAM for use by subsequent computation stages.

\subsection{Non-Linear Engine (NLE)} \label{sec:nle}

The NLE handles the element-wise nonlinear functions that appear throughout NF coupling layers and transformer blocks. Hardware-intensive functions, such as exponential, GeLU, natural logarithm, and reciprocal, are implemented as LUTs, enabling 
single-cycle evaluation without iterative approximation logic. Simpler operations are realized with minimal hardware: ReLU is implemented as a sign-bit-controlled multiplexer that selects between the input and zero, while the max operation uses a dedicated reduction tree.

Composite functions are assembled by chaining these primitives in sequence, entirely within the NLE. Softmax, for example, follows the formulation $\frac{e^x}{\sum_{i=1}^{N} e^x}$ \cite{transpim}, where inputs are first passed through the exponential LUT. The resulting values are then summed via a reduction tree. The reciprocal of this sum is obtained from the reciprocal LUT, and each exponential output is finally scaled by this reciprocal to produce the normalized softmax result. By composing these hardware primitives entirely within the NLE, M3D-FLOW avoids routing intermediate values back through the ME or VAE, minimizing data movement and keeping nonlinear evaluation off the critical compute path.

\section{Dataflow Optimizations and Pipelining} \label{sec:optimization}
M3D-FLOW intelligently maps NF workloads onto its hybrid-bonded M3D PNM substrate to sustain high throughput NF processing under memory and thermal constraints. We first describe how PEs are dynamically partitioned and mapped to NF flow stages, then present the dataflow optimizations that overlap DRAM access with computation, and finally describe the execution pipelining mechanism that converts stage-wise sequential computation into steady-state parallel throughput.

\subsection{Dynamic Partitioning of Processing Engines} \label{sec:sets}
Dynamic partitioning of PEs into logical \textit{sets} allows M3D-FLOW to align hardware parallelism directly with the stage-wise structure of NF models. For a logic die with $N$ PEs executing an NF model with $F$ flow stages, the architecture forms $F$ sets of $N/F$ PEs each. Because sets are logical rather than physical groupings, the partitioning can be reconfigured in software to match different NF architectures and model sizes without any hardware modification.

The mapping from flow stages to sets is one-to-one: each set is assigned exclusively to a single flow stage and stores that stage's parameters in its associated DRAM banks. Since each flow stage depends only on the output of the previous stage, but otherwise executes independently, this assignment creates a natural pipeline boundary between sets that M3D-FLOW exploits, as discussed later in Section~\ref{sec:pipelining}.

As illustrated in Fig.~\ref{fig:architecture}(c), a system with 16 PEs running a four-stage NF model partitions into four sets of four PEs each, with every set performing the same computation type but on stage-specific parameters. To further reduce data access latency, model weights are placed in DRAM tiers according to access frequency: parameters accessed most often are mapped to lower-latency tiers in the tiered M3D memory organization, minimizing average memory access time and ensuring that deeper pipelines do not introduce a memory bottleneck. This tiered memory organization is an inherent property of the underlying M3D DRAM architecture, not a contribution of M3D-FLOW. Prior work \cite{stratum} has exploited this same latency-based tiering for mapping frequently activated mixture of experts to lower-latency tiers, whereas M3D-FLOW applies this physical tiering to NF-specific access patterns, mapping weight tensors by stage-level access frequency.

\subsection{Dataflow Optimization}\label{sec:dataflow}
M3D-FLOW's dataflow is organized around a key principle: memory access and computation must never wait on each other. Before execution begins, model parameters are loaded into the DRAM banks offline via direct memory access (DMA), so that each bank can supply its corresponding PE with preloaded weights from the first cycle of inference. An SRAM buffer between each bank and its PE serves as a staging area that decouples DRAM access from PE computation, absorbing the mismatch between DRAM access latency and the much lower compute latency.

This decoupling is essential because M3D memory access latency exceeds computation latency, which is a consequence of the high parallelism within each PE. To prevent DRAM from becoming a bottleneck, M3D-FLOW employs a prefetching strategy that keeps computation running continuously even during DRAM unavailability. Specifically, during $t_\text{RCD}$ and $t_\text{RP}$ intervals when the HB I/O cannot supply new data, the ME and VAE load operands already present in the SRAM into their own local buffers and proceed with computation. Once new data arrives from DRAM 
and the SRAM begins updating, the ME and VAE continue operating from these local buffers without stalling. This double-buffered overlap of DRAM access and computation effectively hides memory latency, eliminates idle cycles, and ensures that M3D-FLOW operates in a bandwidth-saturated rather than latency-limited regime. When a model's weights exceed SRAM capacity, this same prefetching mechanism overlaps computation with DRAM fetches under an output-stationary dataflow, replacing consumed weights as needed. The cycle-accurate simulator tracks every such fetch round, so reported execution times for large models (e.g., TarFlow \cite{tarflow}) include this full memory overhead.

\subsection{Execution Pipelining} \label{sec:pipelining}
Execution pipelining across sets converts the inherently sequential nature of NF flow stages into system-level parallelism. NF models have a structural property that M3D-FLOW directly exploits for throughput: their flow stages are ordered and parameter-distinct, yet once a stage receives its input, its computation is independent of the others. M3D-FLOW maps this structure onto the logic die by pipelining execution across the sets defined in Section~\ref{sec:sets}, keeping all sets active simultaneously once the pipeline reaches full occupancy. Each set holds the parameters 
of its assigned stage and performs the same computation type as every other set, but on different data. This forms a spatial pipeline that turns stage-level sequentiality into steady-state parallel throughput.

\begin{figure}[tb]
    \centering
    \includegraphics[width=\columnwidth]{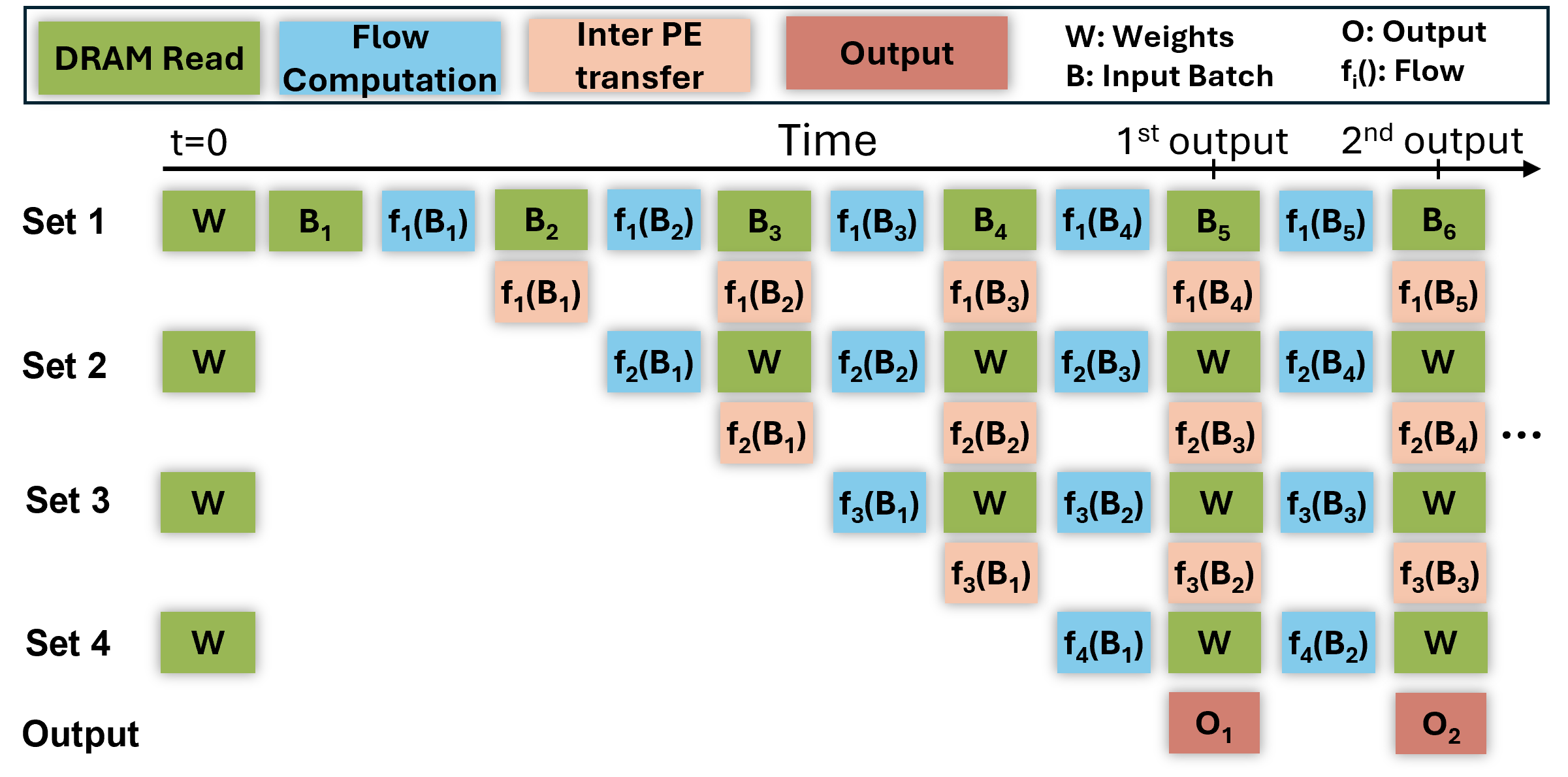}
    \caption{Pipelined execution example for a four-stage normalizing 
    flow model. $f_i(\cdot)$ represents the $i^\text{th}$ flow stage 
    and $O_i$ denotes the final output produced after batch $B_i$ 
    passes through all four sets.}
    \label{fig:pipelining}
\end{figure}

Fig.~\ref{fig:pipelining} illustrates this with a four-stage NF model and $B_S$ input batches mapped onto four sets. At startup, all sets load their respective weight parameters from DRAM into local SRAM, with reads overlapped with computation as described in Section~\ref{sec:dataflow}. Set~1 then processes the first input batch $B_1$ using the ME, VAE, and NLE, while the remaining sets remain idle. Once Set~1 completes, the intermediate output $f_1(B_1)$ is forwarded to Set~2 via the inter-PE communication channel; simultaneously, Set~1 begins processing the next 
batch $B_2$. Set~2 applies its own parameters to $f_1(B_1)$ to produce $f_2(B_1)$, and this pattern propagates across all sets ($f_3 \rightarrow f_4$), as shown in Fig.~\ref{fig:pipelining}.

Once $B_1$ has passed through all four sets, the final output $O_1 = f_4 \cdot f_3 \cdot f_2 \cdot f_1(B_1)$ is produced. Thereafter, a new output is generated after every flow computation cycle as the pipeline runs at full occupancy, so that the number of concurrently active sets translates almost linearly into an increase in inference throughput.

\section{Evaluation} \label{sec:results}
\subsection{Evaluation Setup}
\textbf{System Configuration.} 
M3D-FLOW is evaluated on a monolithic 3D-stackable DRAM with horizontal 1T1C cells, vertical bitlines, and 1024 layers grouped into eight tiers for tiered-latency access \cite{stratum}. The M3D DRAM is synthesized at a 35~nm feature size; detailed parameters are listed in Table~\ref{tab:dram_design} \cite{hsu2025monolithic,stratum}. High-voltage DRAM circuits are placed beneath the array using a 32~nm CuA process, while memory controllers are integrated into a 7~nm logic die. The two 
dies are hybrid-bonded using 1~$\mu$m Cu--Cu pads. Multiplexers bridge the pitch mismatch between the 70~nm bitlines and the bonding pads, enabling burst reads that reduce DRAM read latency. This configuration delivers a total bandwidth of 2.38--3.79~TB/s, depending on the accessed M3D DRAM tier.

The logic die is synthesized using a 7~nm process with the ASAP7 design kit \cite{clark2016asap7} and SiliconCompiler \cite{siliconcompiler}, operating at 1~GHz. Area, latency, and energy are extracted directly from synthesis results and fed into HotSpot \cite{han2021thermal} for 3D thermal modeling (Section~\ref{sec:thermal_analysis}). This is consistent with the methodology used in prior hybrid-bonded 3D-stacked PNM designs \cite{3dpath,stratum,li2026hardware} with architecturally similar configurations. SRAM characteristics are also obtained from the SiliconCompiler \cite{siliconcompiler}. Each PE implements bfloat16 arithmetic via the hybrid binary--stochastic approach described in Section~\ref{sec:me}, and is connected to adjacent PEs through a 128~GB/s inter-PE link (prior works, e.g., \cite{stratum}, commonly use 128~GB/s inter-PE links). Full logic die specifications are provided in Table~\ref{tab:logic_design}.

\begin{table}
\centering
\caption{Configuration Parameters for M3D DRAM. BL: Bitline and WL: Wordline}
\label{tab:dram_design}
\begin{tabular}{|c|c|c|}
\hline
\multicolumn{2}{|c|}{\textbf{Parameters}} & \textbf{Values}\\ \hline
\multirow{7}{*}{\rotatebox[origin=c]{90}{\textbf{M3D}}}
    & Layers & 1024 \\ \cline{2-3}
    & BL Pitch & 70 nm \\ \cline{2-3}
    & WL Pitch & 1$\mu$m \\ \cline{2-3}
    & Staircase Pitch & 500 nm \\ \cline{2-3}
    & Bank Area & 0.439 mm$^{2}$ \\ \cline{2-3}
    & Chip Area & 121 mm$^{2}$ \\ \cline{2-3}
    & Energy per bit & 0.429 pJ \\ \hline
\multirow{8}{*}{\rotatebox[origin=c]{90}{\textbf{DRAM}}}
    & No of wordlines per MAT & 1024 \\ \cline{2-3}
    & No of bitlines per MAT & 1024 \\ \cline{2-3}
    & No of MATs per subarray  & 32 \\ \cline{2-3}
    & No of subarray per bank  & 32 \\ \cline{2-3}
    & No of banks per channel & 16 \\ \cline{2-3}
    & No of channels & 16 \\ \cline{2-3}
    & No of tiers & 8 \\ \cline{2-3}
    & Total Capacity & 32 GB \\ \hline
\multirow{4}{*}{\rotatebox[origin=c]{90}{\textbf{Timing}}}
    & \multirow{2}{*}{$t_{RCD}$ per tier} & [2.29,3.92,5.99,8.50,11.44, \\
    & & 14.82,18.63,22.88] ns\\ \cline{2-3}
    & $t_{RAS}$ & $t_{RCD}$+27.50 ns \\ \cline{2-3}
    & $t_{RP}$ & 4.77 ns \\ \cline{2-3}
    & Total Bandwidth & 2.38 TB/s to 3.79 TB/s \\ \hline
\end{tabular}
\end{table}

\begin{table}
\centering
\caption{Configuration of the Logic Die and Processing Engine (PE). B-to-S: Binary-to-Stochastic, S-to-B: Stochastic-to-Binary.}
\label{tab:logic_design}
\begin{tabular}{|c|c|c|}
\hline
\multicolumn{2}{|c|}{\textbf{Component}} & \textbf{Details}\\ \hline
\multirow{7}{*}{\rotatebox[origin=c]{90}{\textbf{Logic Die}}}
    & Operating Voltage & 0.7 V \\ \cline{2-3}
    & Operating Frequency & 1 GHz \\ \cline{2-3}
    & Data Format & bfloat16 \\ \cline{2-3}
    & No of PEs  & 256 \\ \cline{2-3}
    & Area  & 121 mm$^2$ \\ \cline{2-3}
    & Peak Performance & 192 TFLOPS \\ \cline{2-3}
    & Peak Power  & 38 W \\ \hline
\multirow{7}{*}{\rotatebox[origin=c]{90}{\textbf{PE}}}
    & No of MAC Units in ME (K) & 24 \\ \cline{2-3}
    & No of MACs per MAC unit & 16 \\ \cline{2-3}
    & No of B-to-S and S-to-B units (N) & 96 \\ \cline{2-3}
    & No of adders in VAE & 384 \\ \cline{2-3}
    & SRAM Memory & 192 KB \\ \cline{2-3}
    & Area & 0.309 mm$^2$ \\ \cline{2-3}
    & Inter-PE bandwidth & 128 GB/s \\ \hline
\end{tabular}
\end{table}

\textbf{Baselines and Modeling Methodology.}
M3D-FLOW is compared against seven baselines: AMD EPYC CPU \cite{amd2024epyc5}, NVIDIA A100 \cite{choquette2021nvidia}, TPUv7 \cite{google2024tpu}, Stratum \cite{stratum}, PNM-DDR \cite{h2llm}, 2DRAM-PNM \cite{cao2026two}, and M3D-FLOW-LUT. Since there are no dedicated NF accelerators, we adapt existing designs to run NF workloads. \textit{Stratum} \cite{stratum} is the strongest prior M3D PNM accelerator; it is 
reconfigured to satisfy the pitch-matching constraint of Section~\ref{sec:dram_arch} for a fair comparison. \textit{PNM-DDR} uses Stratum's logic architecture paired with H2-LLM's LPDDR5 DRAM (bank area 6.76~mm$^2$ \cite{h2llm}), constrained to the same 121~mm$^2$ die footprint, resulting in 32 banks of 128~MB each (4~GB total) but 8{,}192 MAC units per PE. \textit{2DRAM-PNM}\cite{cao2026two} stacks two DRAM dies above a logic die, connected via HB and TSV chains. \textit{NeuPIMs} \cite{heo2024neupims} is a PIM accelerator that uses a co-execution scheme, combining a neural processing unit with PIM-based units for matrix and vector operations, respectively. \textit{M3D-FLOW-LUT} is a variant of our design that replaces the BPC encoder with a Bresenham-based LUT converter, enabling variable bitstream lengths at the cost of higher area per converter. Finally, \textit{AMD EPYC CPU} \cite{amd2024epyc5}, \textit{NVIDIA A100} \cite{choquette2021nvidia}, and \textit{TPUv7} \cite{google2024tpu} serve as CPU, GPU, and TPU baselines, respectively. All accelerators are evaluated using an in-house cycle-accurate Python simulator that models five NF workloads --- NICE \cite{nice}, RealNVP \cite{realnvp}, Glow \cite{glow}, MAF \cite{maf}, and TarFlow \cite{tarflow} --- on CIFAR-10 and ImageNet32 (32$\times$32) at a batch size of 128. The simulator tracks data movement and computation across DRAM banks to derive total execution time and energy for each configuration.

\subsection{Area and Power Characterization} \label{sec:area_power_thermal}
\textbf{Area.} 
The logic die is sized to match the 121~mm$^2$ footprint of the M3D DRAM die, which is a prerequisite for full-area hybrid bonding, and comparable to the die area of HBM3 \cite{park2022hbm3}. As shown in 
Table~\ref{tab:area_power_breakdown}, the PHY module consumes 19.8\% of the logic die area, TSVs for power delivery add a negligible 0.1\% overhead (with 2:1 redundancy), and low-voltage memory controllers occupy a further 12.2\%. The remaining 67.9\% is allocated to the 256 PEs, reflecting the compute-centric nature of the design.

Within each PE, SRAM is the dominant component at 56.3\%, reflecting the buffer-heavy prefetch strategy. The ME accounts for 30.5\% of PE area, encompassing all MAC units and binary--stochastic conversion circuits. Both M3D-FLOW and M3D-FLOW-LUT occupy the same total PE area, but differ in the number of B-to-S/S-to-B units accommodated within that budget. VAE occupies 11.5\%; and NLE consumes only 0.3\%, reflecting its LUT-based simplicity. A small residual of 1.2 \% is reserved for HB I/O pads.

\begin{table}[t]
    \centering
    \caption{Area and Power Breakdown of the Logic Die and Processing Engines.}
    \begin{tabular}{|c|c|c|}
    \hline
         \multicolumn{3}{|c|}{\textbf{Logic Die}} \\ \hline
         \textbf{Component} & \textbf{Area (mm$^2$)} & \textbf{Power (W)} \\ \hline
         PHY Module & 23.94 & \multirow{3}{*}{6.79} \\ \cline{1-2}
         Through-Silicon Via & 0.1 &  \\ \cline{1-2}
         Low-Voltage Peripherals & 14.8 &  \\ \hline
         Processing Engines & 79.1 & 31.03 \\ \hline
         \multicolumn{3}{|c|}{\textbf{Processing Engine}} \\ \hline
         \textbf{Component} & \textbf{Area (mm$^2$)} & \textbf{Power (mW)} \\ \hline
         Multiplier Engine & 0.094 & 73.728 \\ \hline
         Vector Adder Engine & 0.035 & 36.05 \\ \hline
         Non-Linear Engine & 0.001 & 1.01 \\ \hline
         SRAM & 0.174 & \multirow{2}{*}{10.425} \\ \cline{1-2}
         Hybrid Bonding I/Os & 0.004 & \\ \hline
    \end{tabular} 
    \label{tab:area_power_breakdown}
\end{table}

\textbf{Power.} 
Power efficiency is critical in hybrid-bonded designs, where the logic die sits directly beneath the DRAM stack. The M3D DRAM die consumes 13~W at peak bandwidth (3.79~TB/s $\times$ 0.429~pJ/bit), while the logic die reaches a peak of 38~W. Within the logic die, PEs account for 82\% of total power, confirming that compute, not data movement, dominates the power budget. At the PE level, the ME consumes 61\% of PE power despite its stochastic design; the VAE contributes 30\%; and the NLE, SRAM, and auxiliary circuits make up the remainder of the PE power. The fact that the ME remains the largest power consumer even with stochastic circuits underscores the importance of the thermal validation in 
Section~\ref{sec:thermal_analysis}.

\subsection{Thermal Validation} \label{sec:thermal_analysis}
The thermal advantage of stochastic computing is not merely theoretical --- it must hold under realistic 3D stacking conditions where the logic die sits directly beneath the DRAM layers and heat has no downward escape path. We verify this through full-stack 3D thermal simulation in HotSpot \cite{han2021thermal}, modeling a vapor chamber heat sink with 75~J/K capacitance, 0.01~K/W resistance, and 1~mm thickness \cite{stratum} at an ambient temperature of 45°C. The considered heat sink is consistent with experimentally validated liquid-cooled microchannel heat sinks operating at 50--350~mW pumping power~\cite{murrieta2024liquid}. M3D-FLOW is modeled with 256 PEs using the power values from Table~\ref{tab:area_power_breakdown} and a DRAM power of 13~W. Stratum is configured with the same bank count, but a higher logic power of 42.67~W \cite{stratum}, reflecting the greater per-MAC energy of its all-binary arithmetic. PNM-DDR uses 
32 banks with a DRAM power of 5.632~W and a logic power of 80~W, 
with that power distributed across a much larger planar bank footprint than either M3D-based design.

\begin{figure}[t]
    \centering
    \includegraphics[width=\columnwidth]{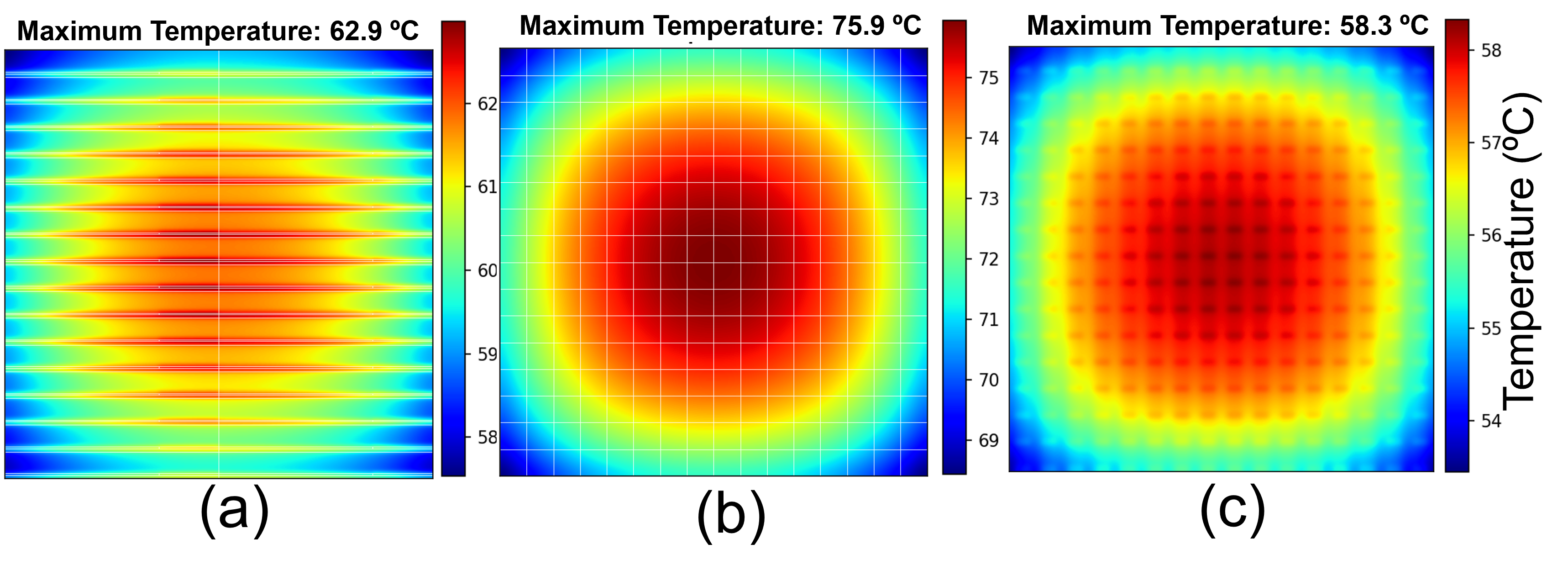}
    \caption{DRAM die thermal maps for (a) PNM-DDR, (b) Stratum, and (c) M3D-FLOW. Temperature scales differ across subfigures, as indicated.}
    \label{fig:thermal_results}
\end{figure}

As shown in Fig.~\ref{fig:thermal_results}, M3D-FLOW achieves the 
lowest peak DRAM temperature of 58.3°C across all three configurations, which is well within the 85°C safe operating limit. This result is especially significant because M3D-FLOW deploys 96K MAC units (50\% more than Stratum's 64K) at the same operating frequency, meaning the lower temperature is achieved with a heavier compute load, not a lighter one. The explanation lies in the power reduction from stochastic mantissa multiplication: replacing full binary multiplier trees with single AND gates reduces per-MAC power sufficiently to offset the increased MAC count, so the logic die's total power remains lower than Stratum's despite running more arithmetic units concurrently. This outcome directly validates the thermal argument of Section~\ref{sec:motivation} and confirms that stochastic computing enables M3D-FLOW to pack greater compute density into the same footprint without exceeding the DRAM thermal ceiling.

The PNM-DDR result provides an instructive contrast that illuminates 
an alternative thermal management strategy. Although PNM-DDR operates at a substantially higher logic power of 80~W, it still remains cooler than Stratum because its 2D monolayer bank layout produces a bank area of 3.38~mm$^2$, which is $7.7\times$ larger than the 0.439~mm$^2$ banks used by M3D-FLOW and Stratum. This larger footprint spreads heat over a much greater surface area, reducing thermal density and preventing hotspot formation even under heavy logic load. Together, the three results confirm that both stochastic power reduction (M3D-FLOW) and physical heat spreading (PNM-DDR) are effective thermal management strategies for hybrid-bonded PNM designs, each well-suited to its respective integration style; M3D-FLOW additionally demonstrates that high compute density and thermal safety can be achieved simultaneously within a compact monolithic 3D stack.

\begin{figure}[t]
    \centering
    \includegraphics[width=\columnwidth]{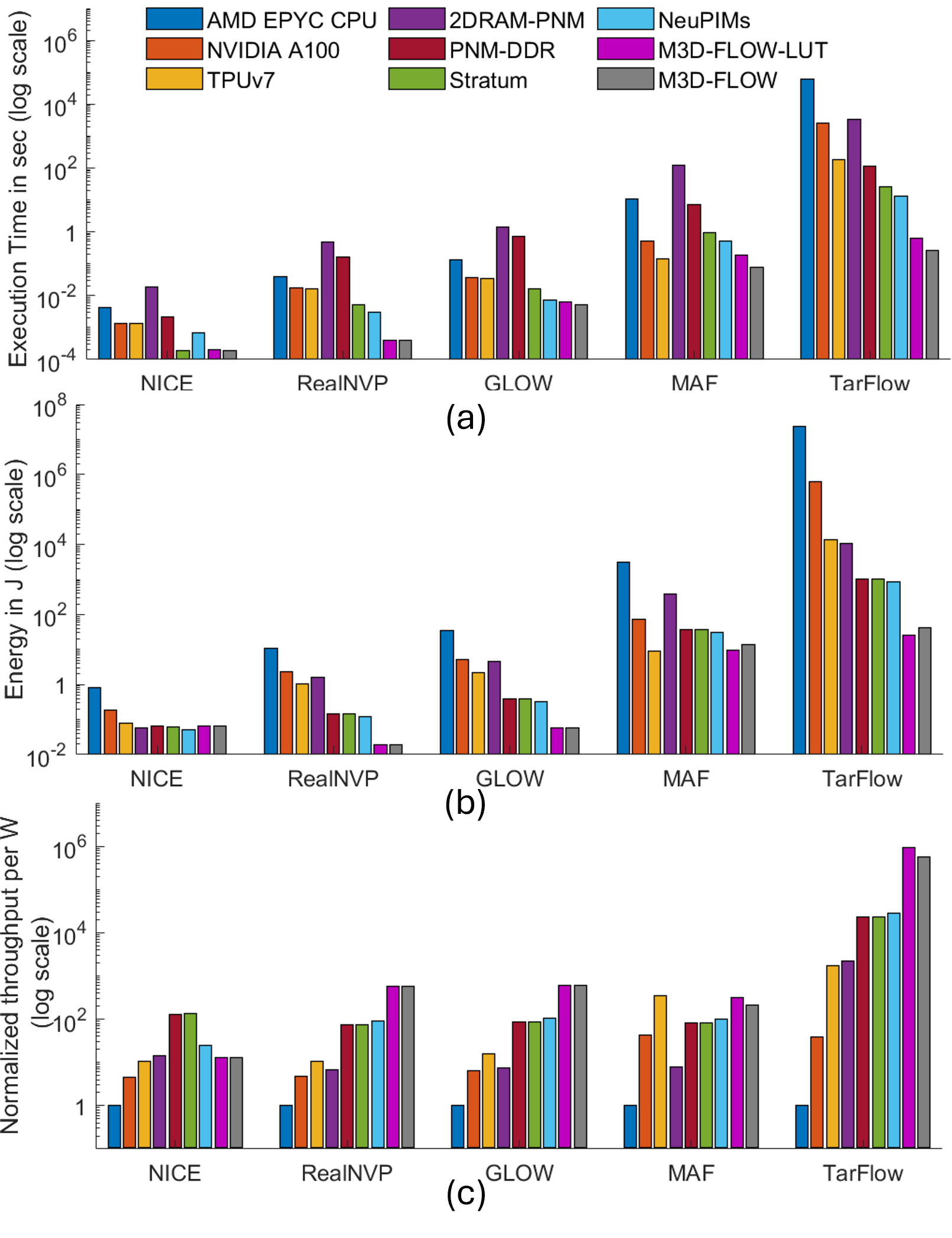}
    \caption{(a) Execution time (seconds), (b) energy consumption (J), and (c) normalized throughput per Watt, reported as geometric means across both datasets for all five NF models at batch size 128.}
    \label{fig:comparison_results}
\end{figure}

\subsection{Performance and Energy Comparison} \label{sec:comparison_baselines}

\textbf{Execution Time.} 
M3D-FLOW achieves the lowest execution time across all five NF models and both datasets. As shown in Fig.~\ref{fig:comparison_results}(a), it reduces execution time by an average of $288\times$, $44\times$, $19\times$, $917\times$, $120\times$, $8.9\times$, $6.6\times$ and $1.5\times$ compared to AMD EPYC CPU, NVIDIA A100, TPUv7, 2DRAM-PNM, PNM-DDR, Stratum, NeuPIMs, and M3D-FLOW-LUT, respectively. These speedups against NVIDIA A100 and TPUv7 do not arise from a memory-capacity advantage (GPUs/TPUs and M3D-FLOW both can store all model parameters in HBM) but from NF-specific workload characteristics and M3D-FLOW's execution model. NF inference is predominantly memory-bandwidth-bound and dominated by affine coupling and elementwise operations that underutilize tensor cores, while the strictly sequential flow-stage structure forces GPUs/TPUs to execute $F$ separate kernels with full HBM round-trips per stage. In contrast, M3D-FLOW keeps parameters and activations near M3D DRAM, and its spatially pipelined PE sets overlap all flow stages in steady state, eliminating per-stage kernel-launch and HBM-traffic overheads and thereby achieving the reported 44$\times$/19$\times$ speedups.  The $8.9\times$ advantage over Stratum reflects M3D-FLOW's higher MAC count (96K vs.\ 64K), dedicated VAE units, and LUT-based NLE, which together reduce per-stage latency. M3D-FLOW outperforms M3D-FLOW-LUT by $1.9\times$ because the larger LUT-based B-to-S converters reduce the number of conversion units per PE, creating an interconversion bottleneck that stalls MAC units.

\textbf{Energy Consumption.} 
The energy advantage of M3D-FLOW is even larger than its throughput 
gains, as shown in Fig.~\ref{fig:comparison_results}(b). M3D-FLOW reduces energy by $895\times$, $76\times$, $14\times$, $34\times$, $5.1\times$, $5\times$, and $3\times$ over AMD EPYC CPU, NVIDIA A100, TPUv7, 2DRAM-PNM, PNM-DDR, Stratum, and NeuPIMs respectively. These reductions are primarily driven by stochastic arithmetic: replacing full binary multiplier trees with single AND gates slashes per-MAC energy, which compounds across 96K concurrent units. Notably, M3D-FLOW-LUT consumes $1.2\times$ less energy than M3D-FLOW, because Bresenham-based LUT converters are more energy-efficient per conversion than BPC circuits. This reveals a direct energy--throughput trade-off between the two converter designs.

\textbf{Throughput per Watt.} Fig. \ref{fig:comparison_results}(c) shows normalized throughput per Watt. M3D-FLOW achieves $896\times$ higher throughput per Watt than the AMD EPYC CPU baseline, compared to $1057\times$ for M3D-FLOW-LUT, $231\times$ for NeuPIMs, $281\times$ for Stratum, $277\times$ for PNM-DDR, $27\times$ for 2DRAM-PNM, $64\times$ for TPUv7, and $11.8\times$ for NVIDIA A100. The combined effect of stochastic power reduction, M3D internal bandwidth, and pipelined execution places M3D-FLOW in a different efficiency class from both conventional accelerators and prior PNM designs. M3D-FLOW-LUT's marginal edge in throughput per Watt over M3D-FLOW reflects its lower energy per conversion, confirming that the choice of the B-to-S converter is the primary efficiency differentiator between the two variants.

\begin{table*}[t]
    \centering
    \caption{Bits-per-Dimension (BPD) across precision configurations. Lower is better. BF16: bfloat16}
    \begin{tabular}{|c|c|c|c|c|c|c|c|c|c|}
    \hline
         \textbf{Model} & \textbf{Dataset} & \textbf{FP32} & \textbf{INT8} & \textbf{BF16} & \textbf{BF16+SC} & \textbf{BF16+LUT$_{256}$} & \textbf{BF16+LUT$_{512}$} & \textbf{BF16+LUT$_{1024}$} & \textbf{BF16+LUT$_{2048}$}  \\ \hline
         NICE & \multirow{4}{*}{CIFAR10} & 5.14 & 5.46 & 5.14 & 5.15 & 5.1412 & 5.1411 & 5.1411 & 5.1411\\ \cline{1-1} \cline{3-10} 
         RealNVP &  & 3.48 & 3.53 & 3.51 & 3.52 & 3.5153 & 3.5147 & 3.5155 & 3.5145 \\ \cline{1-1} \cline{3-10} 
         GLOW &  & 3.39 & 3.70 & 3.66 & 3.87 & 3.6647 & 3.6647 & 3.6645 & 3.6644 \\ \cline{1-1} \cline{3-10} 
         MAF &  & 5.51 & 5.58 & 5.52 & 5.85 & 5.5169 & 5.5165 & 5.5163 & 5.5163 \\ \hline
         TarFlow & AFHQ & 4.63 & 4.77 & 4.64 & 4.59 & 4.64 & 4.63 & 4.63 & 4.63\\ \hline
    \end{tabular}
    \label{tab:bpd_results}
\end{table*}

\subsection{Ablation Study}

\begin{figure}[t]
    \centering
    \includegraphics[width=\columnwidth]{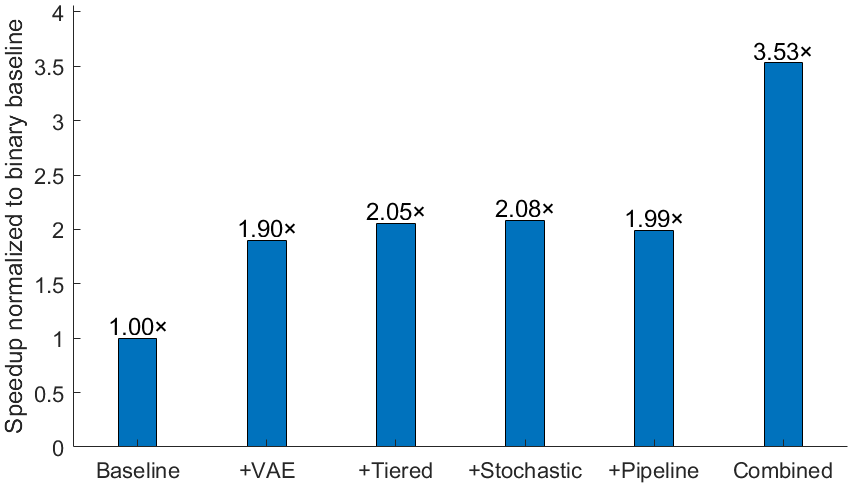}
    \caption{Ablation study comparing the baseline (no optimizations) to designs enhanced with a Vector Adder Engine (VAE), tiered-latency M3D DRAM, stochastic mantissa multiplication, pipelining, and a configuration using all enhancements together.}
    \label{fig:ablation}
\end{figure}

Fig. \ref{fig:ablation} decomposes the M3D-FLOW speedup by enabling each architectural component individually over a baseline that uses a binary logic die without a dedicated VAE, tiered memory, or pipelining. The VAE yields a 1.90$\times$ speedup by offloading element-wise vector operations from the ME's adder tree. Tier-aware memory mapping contributes 2.05$\times$ by exploiting the bandwidth asymmetry between the logic-adjacent and upper DRAM tiers, keeping model parameters at lower tiers where access latency is minimal. Stochastic mantissa multiplication contributes the largest individual gain at 2.08$\times$ by increasing compute density, while pipelined execution contributes 1.99$\times$ by sustaining throughput across sequential flow stages. The combined configuration achieves 3.53$\times$, exceeding every individual component but well below their naive product ($\sim16\times$). This indicates substantial overlap in the bottlenecks each component addresses. Once one component relieves a bottleneck, the marginal benefit of the others shrinks, and pipeline fill/drain overhead at the flow stage boundaries further limits the combined gain. Still, the combined design substantially outperforms any single optimization, confirming the benefits of combining various design enhancement components together instead of employing them individually.

\subsection{Accuracy Impact of Stochastic Computing} \label{sec:error_analysis}
Stochastic mantissa multiplication introduces quantization-like 
approximation error whose magnitude depends on bitstream length. 
M3D-FLOW uses a fixed BPC-based converter with a deterministic 
encoding scheme, which introduces a mean absolute error (MAE) of 
3.9\% in mantissa multiplication. The LUTs in the NLE incur an additional MAE of 0.2\%. This error adds to the precision 
reduction already incurred by using bfloat16 rather than FP32 
training precision. We quantify the combined accuracy impact using bits per dimension (BPD). BPD is chosen over FID because it is derived directly from the model's exact log-likelihood and provides an exact, mathematically grounded measure of how well a normalizing flow models the true data distribution. M3D-FLOW adopts bfloat16 rather than INT8 because INT8 quantization introduces sufficient precision loss to measurably degrade BPD for NF workloads, as confirmed in Table~\ref{tab:bpd_results}. That said, INT8 multipliers present area and power characteristics similar to those of bfloat16 mantissa multipliers, since both operate on comparably sized bit fields in the stochastic domain. An INT8 variant of M3D-FLOW would therefore be equally viable from a thermal and compute-density perspective, preserving the same area-power-thermal trade-offs and performance advantages demonstrated here, with the accuracy-precision trade-off serving as the primary differentiator between the two formats.

As shown in Table~\ref{tab:bpd_results}, M3D-FLOW's stochastic bfloat16 implementation incurs only a 2.2\% BPD increase relative 
to bfloat16 alone, and a 4.1\% increase relative to FP32, yet outperforms INT8 by 0.04\%. This confirms that the energy and area savings of stochastic computing 
come at negligible accuracy cost for NF workloads. Note that absolute BPD values may differ from prior published results due to 
differences in training configuration (epochs, hardware platform).

\begin{table*}[t]
    \centering
    \caption{Per-PE hardware characteristics of multiplier and binary--stochastic interconversion units (B-to-S/S-to-B units) across bitstream lengths and corresponding MAE. Longer bitstreams increase unit area and power, reducing the number of conversion units ($N$) per PE and creating a converter bottleneck, while power density remains consistent across all viable configurations. L: bitstream length; N: B-to-S/S-to-B units per PE; B-to-S: Binary-to-Stochastic; S-to-B: Stochastic-to-Binary.}
    \begin{tabular}{|c|c|c|c|c|c|c|c|}
        \hline
        \multirow{2}{*}{\textbf{L}} & \multicolumn{2}{c|}{\textbf{Multiplier}} & \multicolumn{3}{c|}{\textbf{B-to-S \& S-to-B}} &  \multirow{2}{*}{\textbf{Power Density ($\mu W/\mu m^2$)}} & \multirow{2}{*}{\textbf{MAE (\%)}} \\ \cline{2-6}
         & \textbf{Area ($\mu m^2$)} & \textbf{Power ($\mu W$)} & \textbf{N} & \textbf{Area ($\mu m^2$)} & \textbf{Power ($\mu W$)} & & \\ \hline
         256 & 12 & 14.1 & 40 & 86 & 338.9 & 0.25 & 0.19 \\ \hline
         512 & 27 & 30.7 & 27 & 116.9 & 491.4 & 0.31 & 0.09 \\ \hline
         1024 & 49 & 44.4 & 14 & 175.7 & 524.6 & 0.325 & 0.05 \\ \hline
         2048 & 97 & 54.1 & 5 & 293.4 & 527.8 & 0.324 & 0.02 \\ \hline
         65536 & \multicolumn{6}{c|}{Not viable due to severe area overheads} & 0 \\ \hline
    \end{tabular}
    \label{tab:sensitivity}
\end{table*}

\vspace{5pt}

\begin{figure}[t]
    \centering
    \includegraphics[width=0.80\columnwidth]{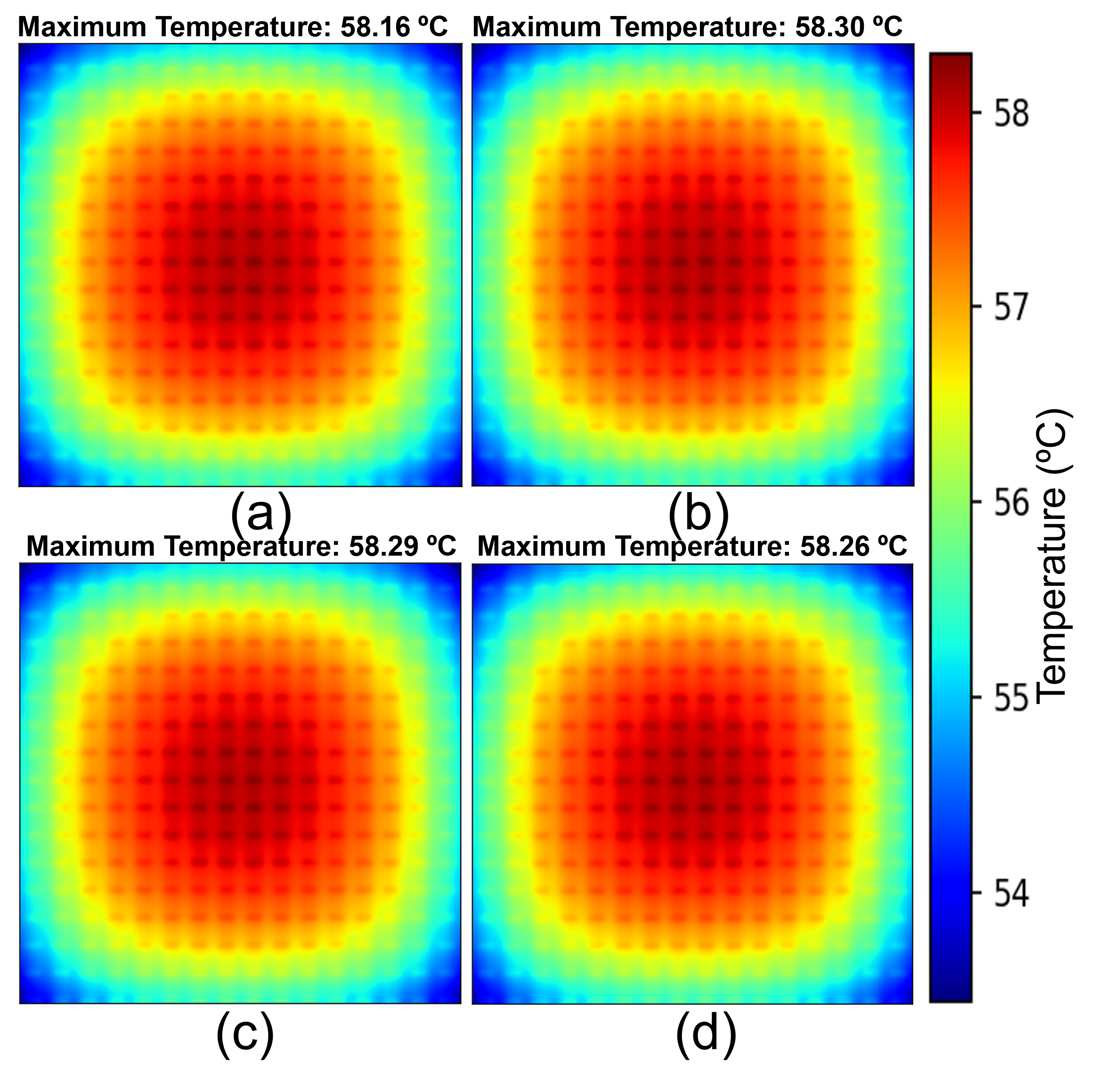}
    \caption{DRAM die thermal maps for (a) M3D-FLOW-LUT$_{256}$, (b) M3D-FLOW-LUT$_{512}$, (c) M3D-FLOW-LUT$_{1024}$ and (d) M3D-FLOW-LUT$_{2048}$.}
    \label{fig:lut_thermal}
\end{figure}

\begin{figure}[t]
    \centering
    \includegraphics[width=\columnwidth]{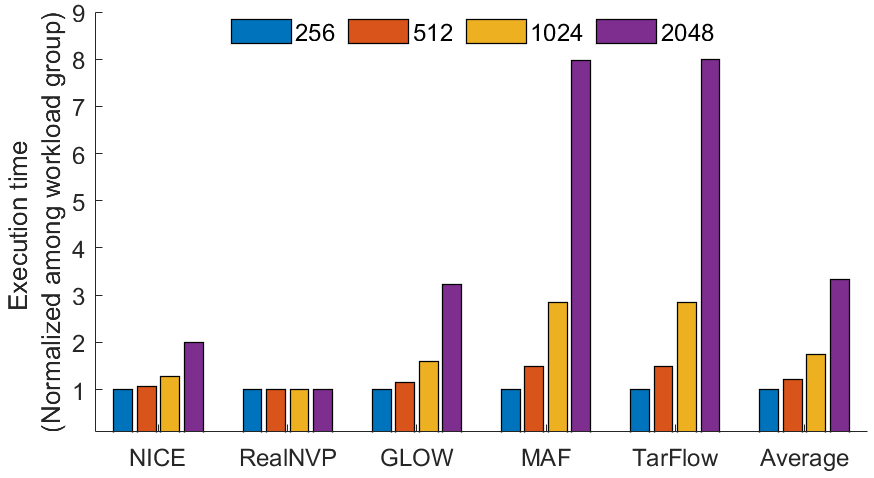}
    \caption{Normalized execution time for M3D-FLOW-LUT variants at bitstream lengths of 256, 512, 1024, and 2048 (batch size of 128). Bars are normalized within each workload group.}
    \label{fig:sensitivity_results}
\end{figure}

\subsection{Bitstream Length vs.\ Area-Latency-Accuracy Trade-offs} \label{sec:sensitivity}

M3D-FLOW-LUT exposes bitstream length as a tunable design parameter,  enabling a principled exploration of the three-way trade-off between area, execution time, and approximation accuracy. The central tension is straightforward: longer bitstreams reduce multiplication error but increase the area of each multiplier, B-to-S unit, and S-to-B unit. Since the number of MAC units ($K$) is held constant, this area growth forces a reduction in the number of conversion circuits ($N$) within the fixed PE area budget, as quantified in Table~\ref{tab:sensitivity}. Fewer converters per PE means more MAC units stall awaiting conversion, directly degrading throughput. Conversely, shorter bitstreams accommodate more converters per PE and sustain higher MAC utilization, at the cost of greater approximation error. At a bitstream length of 2{,}048, the LUT area is fully counterbalanced by the reduced number of B-to-S and S-to-B units that can be accommodated within the fixed area budget, reaching the practical limit of this trade-off. Beyond 2{,}048 bits, further increases in bitstream length are infeasible without reducing the number of multipliers themselves, which would directly cut the peak MAC throughput of the PE. Crucially, because power density 
remains consistent across all configurations (0.25--0.33~$\mu$W/$\mu$m$^2$, Table~\ref{tab:sensitivity}), operating temperature is largely unaffected by bitstream length, as confirmed by the thermal maps in Fig.~\ref{fig:lut_thermal}. This means bitstream length can be tuned freely for performance or accuracy without incurring any additional thermal risk.

The quantitative trade-offs across the four configurations are summarized in Table~\ref{tab:sensitivity} and Fig.~\ref{fig:sensitivity_results}. Increasing bitstream length from 256 to 2{,}048 increases execution time by $3.3\times$, while MAE falls from 0.19 to 0.02, which indicates a $4.7\times$ accuracy improvement at a $3.3\times$ latency cost. Intermediate configurations offer well-defined operating points: the 512-bit and 1{,}024-bit variants incur $1.2\times$ and $1.7\times$ higher execution time than the 256-bit baseline, respectively, while reducing MAE by 52.6\% and 73.7\%. Notably, BPD values remain largely unchanged across all configurations (Table~\ref{tab:bpd_results}), because stochastic approximation errors affect individual mantissa multiplications independently and symmetrically. Therefore, errors tend to cancel rather than accumulate as activations propagate through the successive affine coupling layers of the NF model, where each stage aggregates hundreds to thousands of multiplications before passing its output onward. As a result, per-operation MAE does not compound into model-level output degradation for NF workloads. Although error theoretically vanishes at a bitstream length of 65{,}536, this configuration is impractical: the area required per multiplier would necessitate a severe reduction in both conversion circuits and multiplier count, causing catastrophic throughput loss that far outweighs any accuracy benefit.

These results yield a concrete design guide for practitioners. Latency-critical deployments should select the 256-bit configuration for its $3.3\times$ lower execution time, accepting an MAE of 0.19 that has negligible impact on BPD. Accuracy-sensitive applications tolerating a $1.7\times$ execution time overhead should choose the 1{,}024-bit configuration, which achieves an MAE of 0.05 with no measurable temperature increase. In both cases, the thermal budget remains intact, making bitstream length the primary lever for balancing the latency--accuracy operating point in M3D-FLOW-LUT deployments.

\section{Conclusion}
In this paper, we presented M3D-FLOW, a hybrid-bonded monolithic 3D-stackable processing-near-DRAM architecture tailored to normalizing flow workloads. By tightly integrating logic with M3D DRAM and combining hybrid binary–stochastic computation with pipelined execution 
across dynamically partitioned processing elements, M3D-FLOW simultaneously alleviates memory bottlenecks and improves compute efficiency. Experimental results show that M3D-FLOW delivers at least $6.6\times$ higher performance and $3\times$ better energy efficiency than prior PNM accelerators, while achieving orders-of-magnitude gains over AMD EPYC CPU, NVIDIA A100 and TPUv7 baselines. Despite relying on stochastic mantissa multiplication, the architecture maintains low accuracy degradation, with only a 4.1\% increase in BPD relative to FP32.

Our stochastic design-space exploration further reveals that bitstream length is an effective knob for trading execution time against approximation error: shorter bitstreams improve execution time by up to $3.3\times$ at the cost of higher MAE, while longer bitstreams reduce 
error with modest latency overhead. Thermal analysis confirms that M3D-FLOW remains within safe DRAM temperature limits under full load, owing to the reduced per-MAC power of stochastic multipliers and the compute-density-aware floorplanning of the hybrid-bonded stack. Taken 
together, these results demonstrate that combining M3D integration, near-memory processing, and stochastic computing provides a practical and thermally safe path to efficiently accelerate modern generative models such as normalizing flows.

\section*{Acknowledgments}
This work was supported in part by National Science Foundation (NSF) grants CSR-2443677, CCF-2450614, and CCF-2450615. The authors also thank the anonymous reviewers for their valuable feedback, which helped improve this paper.

\bibliographystyle{IEEEtran}
\bibliography{references}

\begin{IEEEbiography}[{\includegraphics[width=1in,height=1.25in]{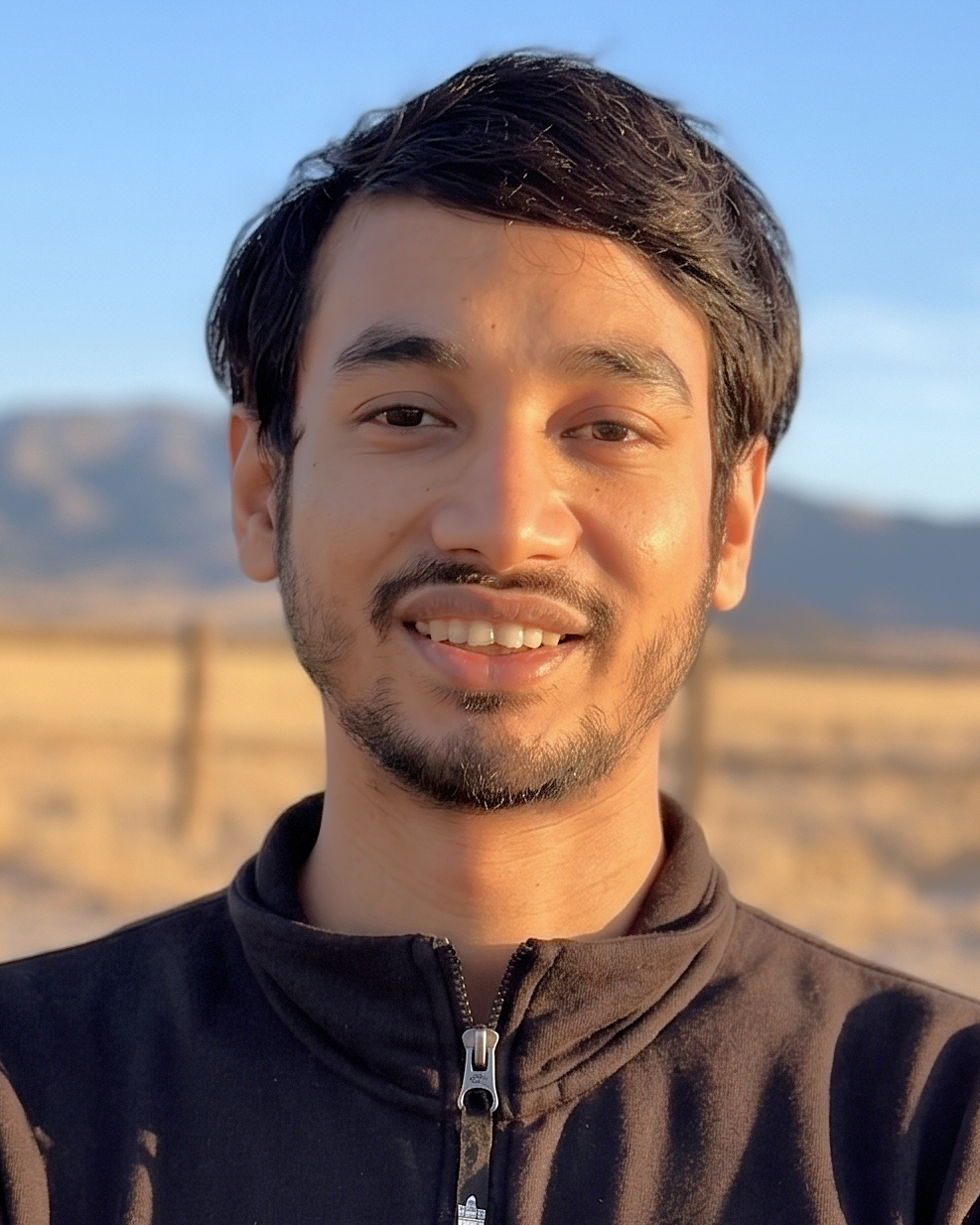}}]{Bipin Thapa Magar}
(Graduate Student Member, IEEE) is currently pursuing the Ph.D. degree in computer engineering with the University of Kentucky, Lexington, KY, USA. His research interests include AI hardware accelerator design, processing-in-memory architectures, and machine learning acceleration using emerging technologies.

\end{IEEEbiography}

\begin{IEEEbiography}[{\includegraphics[width=1in,height=1.25in,clip,keepaspectratio]{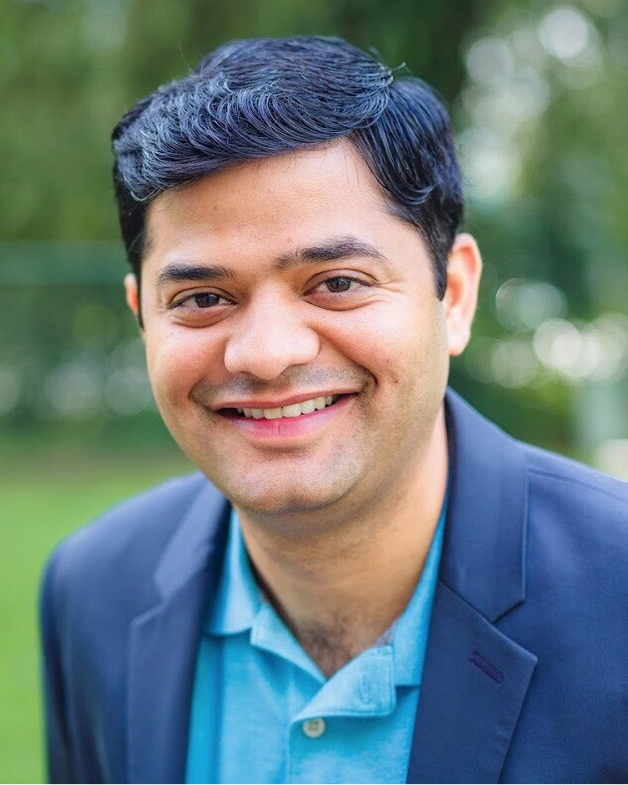}}]{Ishan Thakkar}
(Senior Member, IEEE) received
his Ph.D. degree in Electrical Engineering from
Colorado State University, Fort Collins, CO, USA, in 2018.
He is the Reese S. Terry Associate Professor in the
Department of Electrical and Computer Engineering
at the University of Kentucky, Lexington, KY, USA.
His research interests include computing with unconventional
technologies such as in-memory computing,
integrated electro-photonics, stochastic/unary
computing, and neuromorphic in-materia computing. 
\end{IEEEbiography}

\begin{IEEEbiography}[{\includegraphics[width=1in,height=1.25in,clip,keepaspectratio]{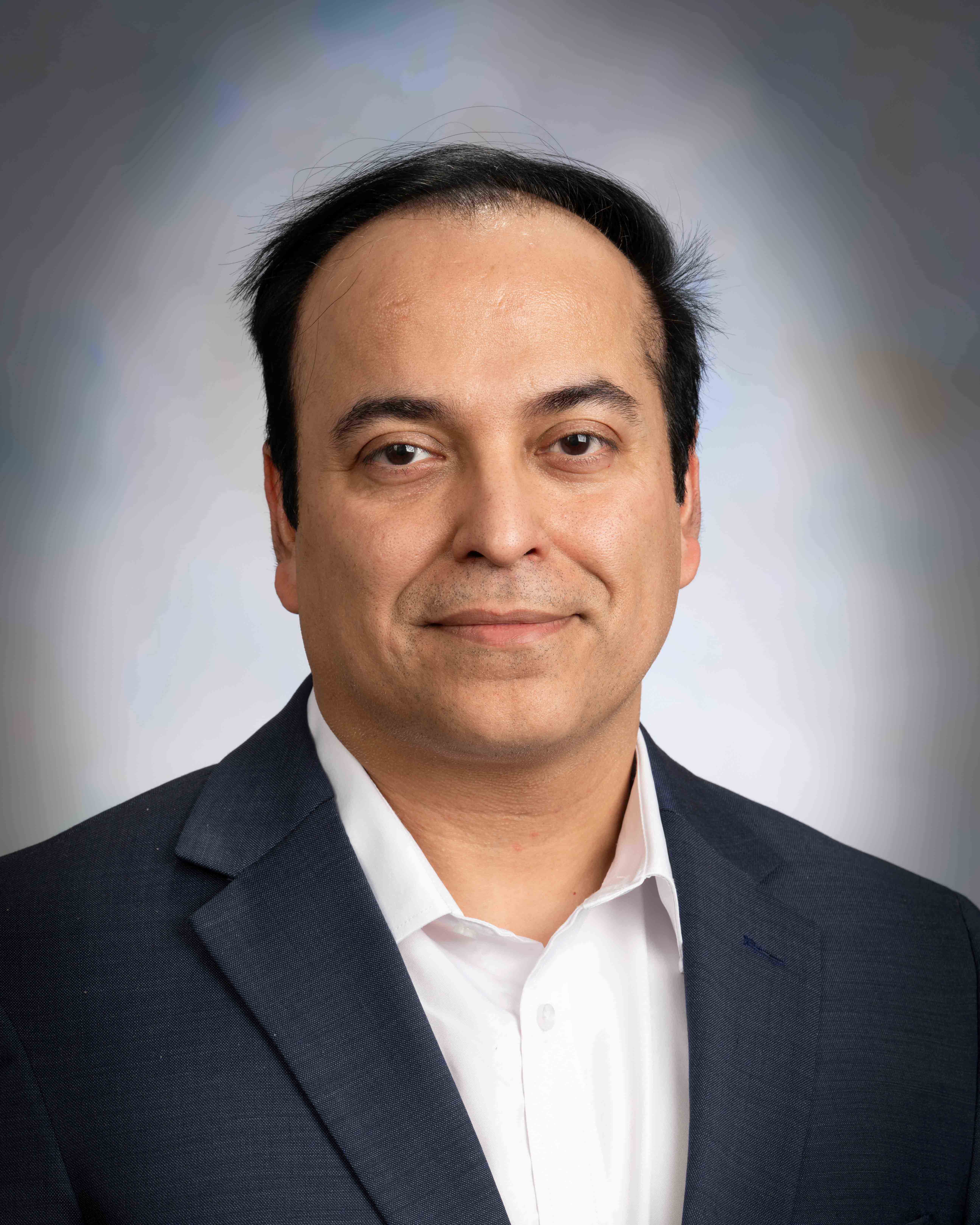}}]{Sudeep Pasricha}
(Fellow, IEEE) received the Ph.D.
degree in computer science from the University of
California, Irvine, CA, USA, in 2008. He is currently the
Aram and Budak Endowed Professor Professor
with the Department of Electrical and Computer
Engineering, Colorado State University, Fort Collins,
CO, USA. His research focuses on the design of
innovative software algorithms, hardware architectures,
and hardware–software co-design techniques
for energy-efficient, fault-tolerant, real-time, and secure
computing. He is a Fellow of the IEEE, a
Fellow of the AIAA, a Distinguished Member of the ACM, and an ACM
Distinguished Speaker.
\end{IEEEbiography}

\vfill

\end{document}